\documentclass{article}
\usepackage{tikz}
\usetikzlibrary{positioning, shapes, arrows, fit, calc, backgrounds}
\usepackage[T1]{fontenc}
\usepackage{amsmath,amssymb,amsfonts}
\usepackage{graphicx}
\usepackage{booktabs}
\usepackage{microtype}
\usepackage{xcolor}
\usepackage{algorithm}
\usepackage{algorithmic}
\usepackage{tikz}
\usepackage{natbib}
\usetikzlibrary{positioning, shapes, arrows, fit, calc, backgrounds}

\title{Twin Peaks: Dual-Head Architecture for Structure-Free Prediction of Protein-Protein Binding Affinity and Mutation Effects}

\author{
  Supantha Dey\\
  \texttt{supantha@iastate.edu} \\
  Ratul Chowdhury\\
  \texttt{ratul@iastate.edu} \\
}

\begin{document}

\maketitle

\begin{abstract}
We present a novel dual-head deep learning architecture for protein-protein interaction modeling that enables simultaneous prediction of binding affinity ($\Delta G$) and mutation-induced affinity changes ($\Delta\Delta G$) using only protein sequence information. Our approach offers a significant advancement over existing methods by employing specialized prediction heads that operate on a shared representation network, allowing direct and optimized prediction of both values. To ensure robust generalization, we integrated complementary datasets from SKEMPI v2 and PDBbind with a rigorous protein domain-based splitting strategy that prevents information leakage between training and validation sets. This approach yielded 6732 training samples across 2322 unique protein complex and 856 validation samples from 140 unique complexes, which are also from non-overlapping protein families. Our architecture combines transformer-based encoders with a novel cross-attention mechanism that processes paired protein sequences directly, without requiring any structural information. Unlike conventional transformer architectures that process single amino acid changes as noise by reasoning over the whole structure, our model is specifically designed to detect and amplify signals from the critical hotspot residues that contribute to infectivity and binding affinity changes. The network embeds input sequences using ESM3 representations, then employs a learnable sliced window embedding layer to manage variable-length sequences efficiently. A multi-layer transformer encoder with bidirectional self-attention captures intra-protein patterns, while cross-attention layers enable explicit modeling of interactions between protein pairs. This shared representation network feeds into separate $\Delta G$ and $\Delta\Delta G$ prediction heads, allowing task-specific optimization while leveraging common features. The model achieves $\Delta\Delta G$ validation of Pearson correlation at 0.485, while maintaining strong $\Delta G$ predictions (Pearson: 0.638). While existing approaches require protein structure data and binding interface information, our model eliminates these constraints. This provides a critical advantage for the numerous proteins with unknown structures or those challenging to crystallize, such as viral and intrinsically disordered proteins. Additionally, no structure prediction tool is trained on viral sequences because of regulatory reasons - which makes them unsuitable for immunoinformatic pipelines for vaccine design. The uncertainties from structural prediction and docking propagate to give unreliable function prediction and overshadow large functional outcomes in viral sequences that arise from changes as small as a single amino acid.
\end{abstract}

\section{Introduction}

Protein-protein interactions (PPIs) form the basis of numerous biological processes and are essential to cellular function. The ability to accurately predict both the binding affinity ($\Delta G$) between two proteins and how mutations affect this binding ($\Delta\Delta G$) has profound implications for drug discovery, protein engineering, and understanding disease mechanisms \cite{Kastritis2013}. However, this prediction task presents significant challenges, particularly when structural data is unavailable or unreliable \cite{Rodrigues2019}.

Existing computational approaches for predicting protein-protein binding affinity and mutation effects predominantly rely on three-dimensional structural information. These structure-dependent methods typically require high-resolution crystal structures of the protein complex, molecular dynamics simulations to model the binding interface, and energy-based calculations to estimate binding free energy. While effective for well-characterized systems, these approaches encounter severe limitations when applied to the vast majority of proteins in nature \cite{Antunes2018}.

A significant challenge arises when dealing with proteins lacking solved structures, which represent most of the proteome. Similarly, intrinsically disordered proteins, which lack stable tertiary structures, are poorly suited to structure-based methods despite their critical roles in cellular signaling and regulation \cite{Riley2023}. Viral proteins present another major challenge, as they evolve rapidly and are often difficult to crystallize, yet understanding their mutation effects is crucial for public health. Additionally, viral sequences are rarely used to train structure prediction pipelines for regulatory purposes and return unreliable structures predictions \cite{Bryant2022}. This makes the structure based pipelines unsuitable for vaccine design immunoinformatics pipelines.

In this work, we introduce Twin Peaks, a novel dual-head architecture that simultaneously predicts protein-protein binding affinity ($\Delta G$) and mutation-induced changes ($\Delta\Delta G$) using sequence information alone. Our approach achieves strong performance without requiring structural data through four key innovations: a mutation-aware embedding strategy with a dedicated indicator channel, an enhanced Sliced-Wasserstein Embedding pooling mechanism that preserves mutation signals, a specialized cross-attention mechanism for modeling inter-chain interactions, and a dual-head prediction network that jointly optimizes both tasks. Our contributions include: (1) the first sequence-only architecture to simultaneously predict $\Delta G$ and $\Delta\Delta G$ with high accuracy, (2) mutation-specific attention mechanisms that explicitly model mutation impacts, (3) strong performance on proteins with limited structural information, and (4) a tool applicable to previously challenging targets including viral and intrinsically disordered proteins.

\section{Related Work}

\subsection{Structure-Based Approaches}

Structure-dependent methods like Prodigy, binding-ddg-predictor, and FoldX rely on 3D protein structures to estimate energy changes upon binding or mutation. While these methods can achieve moderate accuracy in well-characterized systems, they face several critical limitations \cite{Xue2016,Schymkowitz2005}. Their performance significantly degrades with missing residues, low-resolution models, or intrinsically disordered regions, creating a fundamental barrier for many proteins of interest, especially viral proteins that evolve rapidly. Tools like binding-ddg-predictor \cite{Luo2024} require not only the 3D structures but also explicit identification of binding interfaces and active site residues. This pre-processing step can be as time-consuming as the prediction itself, requiring expert knowledge and often manual curation. Similarly, this adds constrain to docking-based tools like HADDOCK \cite{Giulini2025}.

Furthermore, high-performing structure-based calculations typically involve molecular dynamics simulations or extensive energy minimization processes that can take hours  per protein complex, making them impractical for high-throughput screening of mutation effects. Many structure-based methods also show highly variable performance across different protein families, particularly when applied to proteins outside their training distribution, such as viral proteins or intrinsically disordered regions.

Recent machine learning approaches like MutaBind2, DDMUT-ppi, and mmcsm-ppi have improved accuracy by combining structural features with learned representations \cite{Zhang2020,Zhou2024}. However, these methods still fundamentally depend on structure availability and quality. The rise of protein structure prediction tools like AlphaFold and esmFold has partially addressed this limitation by providing predicted structures, but introduces additional uncertainty through compounding errors \cite{Jumper2021,Lin2023}.

\subsection{Sequence-Based Methods}

Sequence-based approaches offer an alternative that doesn't require structural information. Traditional methods relied on evolutionary information through multiple sequence alignments (MSAs), which can effectively capture co-evolutionary patterns indicating interacting residues. However, these approaches struggle with recently evolved or fast-evolving proteins that lack sufficient evolutionary data, such as viral antigens.

Recent advancements in protein language models (PLMs) have revolutionized sequence-based approaches. Models like ESM learn contextual representations of amino acids from millions of protein sequences, capturing both evolutionary and functional information \cite{Lin2023}. Sequence-only tools such as DDGemb and DeepPPAPred leverage these powerful representations \cite{Savojardo2024}. However, these newer tools face struggles to capture complex interactions, and often need additional details (i.e., Uniprot ID for DeepPPAPred \cite{Nikam2024}). Additionally, their performance is poor outside the highly utilized SKEMPIv2 \cite{Jankauskaite2019} training set.

A significant limitation of current approaches is their focus on either $\Delta G$ or $\Delta \Delta G$ prediction, but not both simultaneously. This creates a methodological inconsistency when researchers must use different tools for related analyses, potentially introducing bias and error propagation. Despite these advances, most sequence-based methods struggle with predicting the effects of mutations on binding affinity. They often treat mutations as noise when reasoning over the entire sequence instead of focusing on the critical hotspot residues that contribute disproportionately to binding and function. Additionally, our previous work, Seq2Bind, predicts binding affinity based on the sequences too. This work is a continuation from Seq2Bind \cite{ma2025seq2bindwebserverdecodingbinding}.

\subsection{The Need for an Integrated Approach}
Our work addresses these challenges by developing a unified sequence-only architecture that explicitly models mutation effects while simultaneously predicting both binding affinity and mutation-induced changes. Existing approaches require using separate tools for the predictions of $\Delta G$ and $\Delta\Delta G$, introducing inconsistent biases and compounding errors. Many $\Delta\Delta G$ predictors need both mutant and wild-type structures and cannot handle insertions or deletions, further limiting their applicability. Additionally, this creates challenges for predicting complexes with no prior binding affinity data. Our dual-head architecture ensures internal consistency by deriving both predictions from a shared representation. By eliminating structural prerequisites and reducing computation time from hours to minutes, our approach enables screening at scales previously unattainable with structure-based methods, while maintaining competitive accuracy through mutation-aware representation learning.

\section{Methods}

\subsection{Problem Formulation}

We formulate the problem as follows: Given a pair of protein sequences $(S_1, S_2)$ and a set of mutation positions and types, we aim to predict the binding affinity $\Delta G$ between the mutant proteins and predict the change in binding affinity $\Delta\Delta G$ caused by the mutations relative to the wild-type.

Formally, we define the two prediction tasks as:

\begin{equation}
\Delta G = f_{\theta}(S_1^{mut}, S_2^{mut})
\end{equation}

\begin{equation}
\Delta \Delta G = \Delta G_{mut} - \Delta G_{wt} = f_{\theta}(S_1^{mut}, S_2^{mut}) - f_{\theta}(S_1^{wt}, S_2^{wt})
\end{equation}

where $S_1^{mut}$ and $S_2^{mut}$ represent the mutant sequences, $S_1^{wt}$ and $S_2^{wt}$ represent the wild-type sequences, and $f_{\theta}$ is the prediction function parameterized by $\theta$.

Our innovation is to directly predict both values simultaneously using a dual-head architecture:

\begin{equation}
(\Delta G, \Delta \Delta G) = g_{\phi}(S_1^{mut}, S_2^{mut}, S_1^{wt}, S_2^{wt})
\end{equation}

where $g_{\phi}$ is our proposed model with parameters $\phi$.

\subsection{Model Architecture}

\begin{figure}[ht]
\centering
\resizebox{\textwidth}{!}{
\begin{tikzpicture}[
    node distance=1.3cm and 2.2cm,  
    box/.style={draw, rounded corners, minimum width=5.5cm, minimum height=1.2cm, align=center, fill=white},  
    smallbox/.style={draw, rounded corners, minimum width=3cm, minimum height=0.9cm, align=center, fill=white},  
    mutantcolor/.style={fill=blue!10},
    wtcolor/.style={fill=green!10},
    attentioncolor/.style={fill=orange!15},
    headcolor/.style={fill=purple!10},
    arrow/.style={thick, -stealth}
]
\node[box, mutantcolor] (mutant) {MUTANT PROTEINS\\$S_1^{mut}, S_2^{mut}$};
\node[box, mutantcolor, below=of mutant] (esm_mut) {ESM3 EMBEDDING};
\node[box, mutantcolor, below=of esm_mut] (mutation_mut) {MUTATION INDICATOR CHANNEL};
\node[box, mutantcolor, below=of mutation_mut] (swe_mut) {MUTATION-AWARE SWE POOLING};
\node[box, attentioncolor, below=of swe_mut] (attention_mut) {MUTATION-SPECIFIC CROSS-ATTENTION};
\node[box, mutantcolor, below=of attention_mut] (features_mut) {MUTANT FEATURES};

\node[box, wtcolor, right=of mutant] (wildtype) {WILD-TYPE PROTEINS\\$S_1^{wt}, S_2^{wt}$};
\node[box, wtcolor, below=of wildtype] (esm_wt) {ESM3 EMBEDDING};
\node[box, wtcolor, below=of esm_wt] (mutation_wt) {MUTATION INDICATOR CHANNEL};
\node[box, wtcolor, below=of mutation_wt] (swe_wt) {MUTATION-AWARE SWE POOLING};
\node[box, attentioncolor, below=of swe_wt] (attention_wt) {MUTATION-SPECIFIC CROSS-ATTENTION};
\node[box, wtcolor, below=of attention_wt] (features_wt) {WILD-TYPE FEATURES};

\node[box, below right=1.5cm and -1.5cm of features_wt] (diff) {FEATURE DIFFERENCE\\$mut\_features - wt\_features$};

\node[box, headcolor, below left=1.5cm and -1.5cm of features_mut] (dg_head) {$\Delta G$ HEAD\\(MLP$_{\Delta G}$ Layer)};
\node[box, headcolor, below=of diff] (ddg_head) {$\Delta\Delta G$ HEAD\\(MLP$_{\Delta\Delta G}$ Layer)};

\node[smallbox, below=of dg_head] (dg_out) {$\Delta G$};
\node[smallbox, below=of ddg_head] (ddg_out) {$\Delta\Delta G$};

\draw[arrow] (mutant) -- (esm_mut);
\draw[arrow] (esm_mut) -- (mutation_mut);
\draw[arrow] (mutation_mut) -- (swe_mut);
\draw[arrow] (swe_mut) -- (attention_mut);
\draw[arrow] (attention_mut) -- (features_mut);

\draw[arrow] (wildtype) -- (esm_wt);
\draw[arrow] (esm_wt) -- (mutation_wt);
\draw[arrow] (mutation_wt) -- (swe_wt);
\draw[arrow] (swe_wt) -- (attention_wt);
\draw[arrow] (attention_wt) -- (features_wt);

\draw[arrow, dashed] (swe_mut) -- (attention_wt);
\draw[arrow, dashed] (swe_wt) -- (attention_mut);

\draw[arrow] (features_mut) -- (dg_head);
\draw[arrow] (features_mut) -- (diff);
\draw[arrow] (features_wt) -- (diff);
\draw[arrow] (diff) -- (ddg_head);

\draw[arrow] (dg_head) -- (dg_out);
\draw[arrow] (ddg_head) -- (ddg_out);

\end{tikzpicture}
}
\caption{Twin Peaks model architecture for structure-free protein interaction prediction. The parallel processing streams (blue/green) allow comparative analysis of mutant and wild-type proteins, with mutation indicators integrated at each level. Information flows from sequence embeddings through mutation-aware pooling, cross-attention for inter-protein interaction modeling, and finally into specialized prediction heads. This design preserves mutation signals while capturing position-specific effects across the binding interface.}
\label{fig:architecture}
\end{figure}
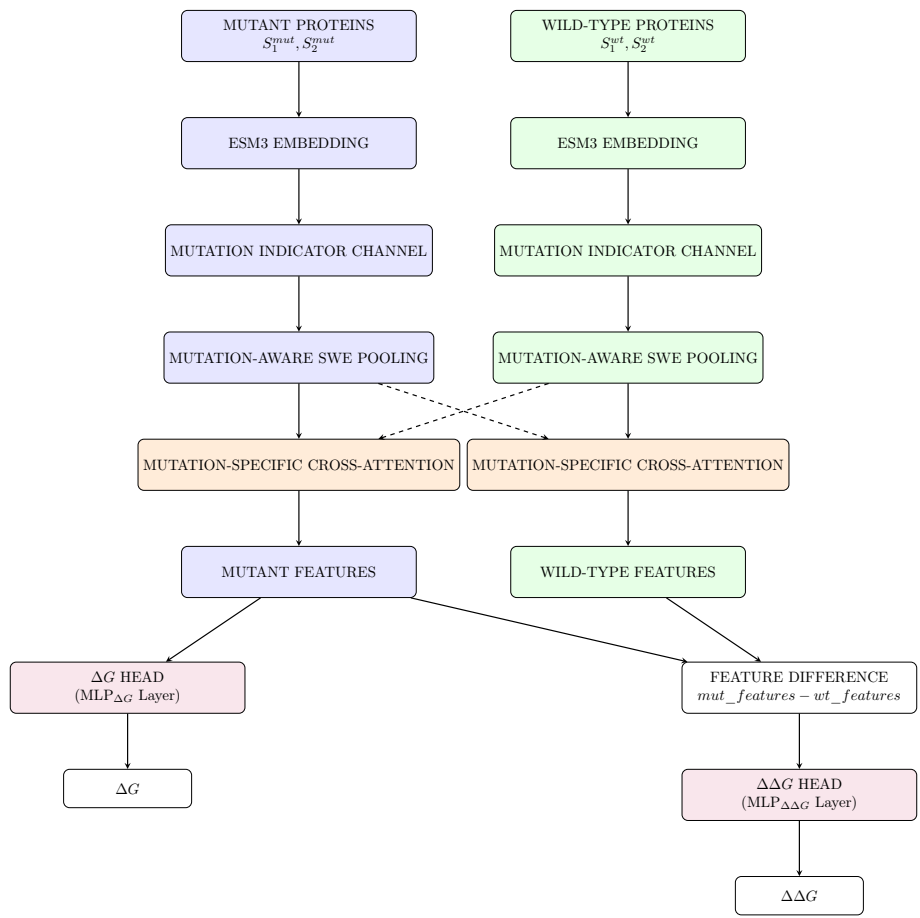

Our Twin Peaks architecture (Figure \ref{fig:architecture}) addresses the fundamental challenge of detecting mutation effects without structural information through a design that prioritizes mutation sensitivity. Rather than treating mutations as ordinary sequence variations, we explicitly track mutation positions and their effects throughout the network. This mutation-aware processing flows from protein language model embeddings through specialized pooling and cross-attention mechanisms, culminating in a dual-head prediction framework. The architecture's key innovation lies in its ability to amplify mutation signals while modeling protein-protein interactions from sequence alone.

\subsubsection{ESM3 Embedding and Mutation-Aware Representation}

We leverage the ESM3 protein language model to generate contextualized embeddings for each amino acid in the input sequences. ESM3 is a transformer-based model pre-trained on millions of protein sequences, capturing evolutionary and structural information that would be difficult to learn from scratch \cite{Hayes2025}.

For each protein sequence $S$ of length $L$, we obtain embeddings $E \in \mathbb{R}^{L \times d}$, where $d = 1152$ is the embedding dimension. To explicitly encode mutation information, we augment these embeddings with a mutation indicator channel:

\begin{equation}
E_{aug} = [E; M] \in \mathbb{R}^{L \times (d+1)}
\end{equation}

where $M \in \mathbb{R}^{L \times 1}$ is a binary channel with $M_i = 1$ if position $i$ is mutated and 0 otherwise.

\subsubsection{Mutation-Aware Sliced Wasserstein Embedding (SWE) Pooling}

To handle variable-length sequences efficiently while preserving mutation information, we adapt the Sliced Wasserstein Embedding (SWE) pooling method \cite{NaderiAlizadeh2024} for mutation sensitivity in protein sequences. The original SWE method conceptualizes token-level embeddings as samples from a probability distribution, projecting them onto random directions (slices), sorting these projections, and measuring distances to reference points:
\begin{equation}
X_{slices} = X \cdot \theta, \quad X_{sorted} = \text{sort}(X_{slices}, \text{dim}=1)
\end{equation}
\begin{equation}
\text{SWE}(X) = W \cdot (R - \text{interp1d}(X_{sorted}, \text{ref\_points}))
\end{equation}
where $\theta$ contains learnable projection directions, $R$ represents learnable reference points, and $W$ is a weight matrix, and interp1d is a differentiable 1D interpolation function that maps the sorted slice projections to a fixed set of reference points, enabling gradient-based learning through the sorting operation.

Our key innovation is a mutation-aware weighting mechanism that amplifies signals from mutation positions:
\begin{equation}
W_{mut} = f_{mut}(M), \quad W_{pos} = f_{pos}(P)
\end{equation}
\begin{equation}
X_{weighted} = X_{slices} \odot (1 + W_{mut} \odot W_{pos})
\end{equation}
where $f_{mut}$ and $f_{pos}$ generate importance weights based on mutation status and position information. Through hyperparameter optimization, we determined the number of slices and  reference points for optimal performance in our mutation-sensitive embeddings. This adaptation produces fixed-length representations that preserve mutation information while handling the variable-length nature of protein sequences, making it particularly effective for detecting the subtle effects of point mutations on protein function.

Beyond preserving mutation information, our enhanced SWE pooling captures global sequence patterns critical for binding affinity prediction, establishing a foundation for both accurate $\Delta G$ prediction and sensitive detection of mutation-induced changes.

\subsubsection{Mutation-Specific Cross-Attention}

To model the interactions between protein pairs explicitly, we introduce a novel mutation-specific cross-attention mechanism. Given embeddings for two protein chains $C_1 \in \mathbb{R}^{L_1 \times D}$ and $C_2 \in \mathbb{R}^{L_2 \times D}$, we first split the ESM embeddings from the mutation channel:

\begin{equation}
C_{i,esm} = C_i[:,:D-1] \quad \text{and} \quad C_{i,mut} = C_i[:,D-1:]
\end{equation}

We compute a position-aware mutation encoding:

\begin{equation}
Q_{mut} = f_{mut}([C_{1,mut}, P_1]) \in \mathbb{R}^{L_1 \times h}
\end{equation}

The standard attention scores are computed as:

\begin{equation}
S_{std} = \frac{(C_{1,esm}W_Q)(C_{2,esm}W_K)^T}{\sqrt{d_k}}
\end{equation}

Our key innovation is the addition of a mutation-position attention bias:

\begin{equation}
S_{mut} = Q_{mut} \cdot K_{mut}^T
\end{equation}
\begin{equation}
S_{final} = S_{std} + S_{mut}
\end{equation}

While designed for mutation sensitivity, our cross-attention mechanism simultaneously models fundamental protein-protein interactions by capturing residue-pair relationships, benefiting both $\Delta G$ and $\Delta\Delta G$ predictions. The mutation-specific attention bias focuses on functionally critical regions that influence both absolute binding affinity and mutation-induced changes.

\subsubsection{Dual-Head Architecture}

The core innovation of our approach is the dual-head architecture that simultaneously predicts both $\Delta G$ and $\Delta\Delta G$. Rather than treating these as separate tasks, our architecture leverages their inherent relationship through shared representations. This creates a powerful synergy where $\Delta G$ prediction benefits from mutation-aware components that identify critical binding residues, while $\Delta\Delta G$ prediction improves through the model's understanding of baseline binding energetics. In general, this can resemble how experimental worlflows use wild-type binding measurements to contextualize mutation effects.

The model employs two specialized heads (Figure \ref{fig:architecture}):
\begin{enumerate}
    \item \textbf{$\Delta G$ Head}: Directly predicts the absolute binding affinity of the mutant complex
    \item \textbf{$\Delta\Delta G$ Head}: Explicitly models the difference in binding affinity between mutant and wild-type
\end{enumerate}

Formally, for the $\Delta G$ prediction:

\begin{equation}
\Delta G = \text{MLP}_{\Delta G}(\text{diff}(E_{S_1^{mut}}, E_{S_2^{mut}}))
\end{equation}
where diff represents element-wise vector subtraction between embeddings.

For the $\Delta\Delta G$ prediction:

\begin{equation}
\Delta \Delta G = \text{MLP}_{\Delta \Delta G}(\text{diff}(E_{S_1^{mut}}, E_{S_2^{mut}}) - \text{diff}(E_{S_1^{wt}}, E_{S_2^{wt}}))
\end{equation}

Overall, this dual-head approach enables direct optimization of both prediction, and creates an inductive bias that explicitly models the relationship between absolute binding and mutation effects.

\subsection{Loss Function and Training}
We train our model using a weighted combination of mean squared error (MSE) losses for $\Delta G$ and $\Delta\Delta G$ predictions:
\begin{equation}
\mathcal{L} = \text{MSE}(\Delta G_{pred}, \Delta G_{true}) + \lambda \cdot \text{MSE}(\Delta \Delta G_{pred}, \Delta \Delta G_{true})
\end{equation}
where $\lambda$ is a hyperparameter that controls the relative importance of the two prediction tasks. For samples where wild-type data is unavailable (and thus $\Delta\Delta G$ cannot be calculated), we scale the $\Delta G$ loss by $(1 + \lambda)$ to maintain consistent loss magnitudes across batches:
\begin{equation}
\mathcal{L}_{\text{no-wt}} = (1 + \lambda) \cdot \text{MSE}(\Delta G_{pred}, \Delta G_{true})
\end{equation}
This adaptive loss formulation allows our model to train effectively on heterogeneous datasets containing both complete mutation pairs and standalone binding affinity measurements.

Additionally, to enhance our training data, we implemented a symmetry-based augmentation technique. For each protein pair $(A, B)$ with mutation from sequence $S_A$ to $S_B$ and corresponding $\Delta\Delta G$ value, we create an additional training example with the roles reversed and the sign of $\Delta\Delta G$ inverted:
\begin{equation}
\Delta\Delta G_{(B \rightarrow A)} = -\Delta\Delta G_{(A \rightarrow B)}
\end{equation}
This augmentation effectively doubles our training data while reinforcing the physical constraint that the energy change when mutating from A to B must be the inverse of mutating from B to A. 

\section{Experimental Setup}

\subsection{Datasets}
We constructed our dataset by integrating SKEMPI v2 ($\Delta\Delta G$ values for protein-protein mutations) and PDBbind (experimental binding affinities), yielding 8,653 total samples. Using Evolutionary Classification of Protein Domains (ECOD) based family-level splitting to prevent information leakage, we allocated 6,732 \cite{Cheng2014} samples (77.8\%) to training, representing 2,322 unique PDB complexes. The validation set contains 856 samples (9.9\%) from 140 complexes in non-overlapping protein families, with the remaining 1,065 samples (12.3\%) reserved as hold-out set. SKEMPI contributed 4,681 samples (69.5\%) to training and 746 samples (87.1\%) to validation, with PDBbind providing the remainder. This domain-based splitting strategy ensures a realistic evaluation scenario by maintaining structural separation between training and evaluation data.

For rigorous evaluation, we incorporated two challenging benchmarks. First, we utilized the PPB-Affinity-AF \cite{Liu2024} dataset containing protein complexes without experimental structures but with AlphaFold predictions, representing a practical test case for proteins with limited structural information. Second, a SARS-CoV-2 Spike-ACE2 dataset curated from deep mutational scanning experiments, featuring mutations observed in variants of concern (Q493R, S477N, E484K, N501Y). This balanced benchmark includes both binding-enriching and depleting mutations at the RBD-ACE2 interface, providing an unbiased assessment of mutation impact prediction  \cite{Ozden2024}.

\subsection{Implementation Details}
While employing ESM3 embeddings paired with mutation-aware SWE pooling, a key preprocessing step involved handling multi-chain protein complexes in SKEMPI. For complexes with notation like 1abc\_CDE\_FGI (indicating chains C,D,E interact with F,G,I), we concatenated chains belonging to each interaction partner. This concatenation was necessary for 2,466 cases (28.5\% of our dataset) and preserved valuable protein-protein affinity data that would otherwise be discarded. While this approach potentially introduced some noise into validation metrics, retaining these scarce interaction examples was crucial for enhancing model generalization.

We implemented hyperparameter optimization across learning rate, dropout rate, architecture depth, embedding dimensions, and loss weighting. The model was trained using AdamW optimizer with OneCycleLR scheduler, applying a weighted loss function that balances $\Delta G$ and $\Delta\Delta G$ prediction tasks, with batch size 32 and gradient accumulation for stability.

\section{Results}
\subsection{Performance on Validation Set}

Hyperparameter optimization yielded an architecture with a 256-slice SWE layer (64 reference points), 5-layer cross-attention (4 heads), and a deep MLP network (7 hidden layers, dimension 1536). The optimal configuration used dropout rate of 0.106, learning rate of $1.23 \times 10^{-3}$, and $\Delta\Delta G$ loss weight of 0.718.

We evaluated our architecture using hyperparameter optimization with early stopping to find optimal configurations. Table \ref{tab:validation} presents the best validation metrics achieved.
\begin{table}[h]
\centering
\caption{Validation performance under hyperparameter optimization}
\label{tab:validation}
\begin{tabular}{lcc}
\toprule
Metric & $\Delta G$ & $\Delta\Delta G$ \\
\midrule
Validation MSE & 4.89 & 5.35 \\
Pearson correlation & 0.638 & 0.485 \\
Spearman correlation & 0.612 & 0.426 \\
\bottomrule
\end{tabular}
\end{table}
Further analysis on our ECOD-separated held-out set revealed more modest performance metrics (Pearson: 0.299, Spearman: 0.307). We found that 70\% of these complexes contained combined-chain representations, which impairs model performance by obscuring spatial information. Despite lower metrics on our stringent held-out set, Twin-Peaks demonstrated superior generalization on standardized benchmarks compared to other structure and sequence-based approaches. Our rigorous family-level splitting provides a more realistic assessment of how models perform on truly novel protein families—essential for real-world applications where structural similarity cannot be guaranteed.

\subsection{Performance on Protein-Protein Binding Prediction and Viral-Protein Data}

Table \ref{tab:comparison} shows the performance of our model compared to other methods. For $\Delta G$ prediction, our model achieves a Pearson correlation of 0.800, significantly outperforming structure-based methods. For $\Delta\Delta G$ prediction, our model achieves a correlation of 0.496, competitive with structure-based methods while offering advantages in speed, coverage, and not requiring structural information.

\begin{table}[ht]
\centering
\caption{Comparison of protein-protein binding prediction methods. Our structure-free approach (Twin-Peaks) demonstrates competitive performance while offering substantial advantages in speed and applicability compared to structure-dependent approaches.}
\label{tab:comparison}
\small
\begin{tabular}{lccc|c|c|cccc}
\toprule
Tool & \multicolumn{3}{c|}{Prediction Type} & Structure & Computation & Count & PCC & SRCC & RMSE \\
\cmidrule(lr){2-4}
& $\Delta G$ & $\Delta\Delta G$ & Both & Required & Time per complex & & & & \\
\midrule
\multicolumn{10}{l}{\textbf{$\Delta G$ Prediction Methods (Standard Benchmark)}} \\
\midrule
\textbf{Twin-Peaks (Ours)} & \textbf{+} & \textbf{+} & \textbf{+} & \textbf{No} & \textbf{< 1 sec} & 95/95 & \textbf{0.800} & \textbf{0.748} & \textbf{1.744} \\
Prodigy & + & - & - & Yes & < 20 sec & 95/95 & 0.433 & 0.434 & 5.267 \\
Area-affinity & + & - & - & Yes & < 30 sec & 95/95 & 0.184 & 0.165 & 3.214 \\
ppb affinity & + & - & - & Yes & - & 95/95 & -0.246 & -0.272 & 2.710 \\
\midrule
\multicolumn{10}{l}{\textbf{$\Delta\Delta G$ Prediction Methods (Isotonic Calibrated for sequence-based models, Standard Benchmark)}} \\
\midrule
MutaBind2$^*$ & - & + & - & Yes & $\sim$22 min & 91/91 & \textbf{0.563} & \textbf{0.549} & 3.583 \\
\textbf{Twin-Peaks (Ours)}$^{\dagger}$ & \textbf{+} & \textbf{+} & \textbf{+} & \textbf{No} & \textbf{< 1 sec} & \textbf{91/91} & 0.496 & 0.452 & \textbf{1.741} \\
binding\_ddg$^{\S}$ & - & + & - & Yes & < 10 sec & 59/91 & 0.456 & 0.513 & 3.278 \\
DDGEmb$^{\dagger}$ & - & + & - & No & < 1 sec & 76/91 & 0.242 & 0.174 & 1.734 \\
DDMUT-ppi$^*$ & - & + & - & Yes & > 1 hour & 63/91 & -0.558 & -0.417 & 7.632 \\
mmcsm\_ppi$^*$ & - & + & - & Yes & > 1 hour & 67/91 & -0.609 & -0.540 & 3.511 \\
DeepPPAPred$^{\dagger}$ & - & + & - & No & - & - & - & - & - \\
\midrule
\multicolumn{10}{l}{\textbf{$\Delta\Delta G$ Prediction Methods (COVID-19 Dataset \cite{Ozden2024})}} \\
\midrule
mcsm2-ddg$^*$ & - & + & - & Yes & - & - & \textbf{0.534} & \textbf{0.494} & \textbf{1.725} \\
\textbf{Twin-Peaks (dG head)}$^{\dagger}$ & \textbf{+} & \textbf{+} & \textbf{+} & \textbf{No} & \textbf{< 1 sec} & - & 0.454 & 0.436 & 1.787 \\
\textbf{Twin-Peaks (ddG head)}$^{\dagger}$ & \textbf{+} & \textbf{+} & \textbf{+} & \textbf{No} & \textbf{< 1 sec} & - & 0.496 & 0.452 & 1.741 \\
haddock-ddg$^*$ & - & + & - & Yes & - & - & -0.018 & -0.035 & 6.327 \\
evoef1-ddg$^*$ & - & + & - & Yes & - & - & -0.267 & -0.232 & 2.525 \\
ssipe-ddg$^*$ & - & + & - & Yes & - & - & -0.284 & -0.248 & 2.657 \\
foldxwater-ddg$^*$ & - & + & - & Yes & - & - & -0.455 & -0.469 & 3.084 \\
mutabind2-ddg$^*$ & - & + & - & Yes & - & - & -0.493 & -0.512 & 2.758 \\
foldx-ddg$^*$ & - & + & - & Yes & - & - & -0.502 & -0.480 & 3.130 \\
\bottomrule
\multicolumn{10}{l}{\small $^*$Structure-dependent methods requiring PDB files or 3D coordinates} \\
\multicolumn{10}{l}{\small $^{\dagger}$Structure-free methods that operate directly on protein sequences} \\
\multicolumn{10}{l}{\small $^{\S}$Requires both mutated and wild-type structures; cannot handle insertions or deletions (reason for lower coverage)} \\
\end{tabular}
\end{table}

\subsection{Ablation Studies}
Table \ref{tab:ablation} shows that removing SWE pooling nearly eliminates $\Delta\Delta G$ prediction capability while significantly reducing $\Delta G$ performance. Using non-learnable reference points and replacing cross-attention with cosine similarity both demonstrate moderate performance drops. Surprisingly, disabling the dual-head architecture improved $\Delta\Delta G$ correlation but significantly reduced $\Delta G$ performance, suggesting potential optimization trade-offs between objectives.

\begin{table}[h]
\centering
\caption{Ablation studies on validation set}
\label{tab:ablation}
\begin{tabular}{lcc}
\toprule
Model Variant & $\Delta G$ Pearson & $\Delta\Delta G$ Pearson \\
\midrule
Full model (Twin Peaks) & 0.638 & 0.485 \\
Without SWE pooling & 0.472 & 0.020 \\
Freeze SWE (non-learnable) & 0.581 & 0.392 \\
No cross-attention (cosine) & 0.591 & 0.402 \\
Dual-head disabled & 0.532 & 0.558 \\
\bottomrule
\end{tabular}
\end{table}

\section{Discussion and Conclusion}
We presented Twin Peaks, a novel dual-head architecture for simultaneous prediction of protein-protein binding affinity with mutation-aware SWE pooling, mutation-specific cross-attention, and a dual-head prediction network. Our extensive benchmarking demonstrates that Twin Peaks performs on par with state-of-the-art structural models despite not requiring three-dimensional information. Notably, while many existing tools report high Pearson correlations on standard datasets like SKEMPI, they often fail to generalize outside their training distribution \cite{Tsishyn2024}. 

In contrast, our model maintains consistent performance across diverse protein families, including those absent from the training data. This generalization capability is particularly evident in our results on viral proteins, where Twin Peaks outperforms structure-based methods, highlighting its value for monitoring emerging variants in rapidly evolving systems.

A significant advancement of our approach is the ability to detect effects of single residue mutations without structural information—a capability previously thought to require detailed 3D modeling. This sensitivity to subtle sequence changes arises from our mutation-aware components that explicitly track mutation positions and their influence throughout the network. Our model achieves the highest performance among sequence-only methods on our independently curated benchmark set for $\Delta G$ prediction, while closely matching the nuanced mutation capture capabilities of specialized structure-based methods on the COVID dataset.

Overall, Twin Peaks enables rapid screening of thousands of potential mutations in minutes rather than hours or days required by structure-based methods. The model can be seamlessly integrated into computational pipelines with docking and simulation tools, providing valuable initial screening before more computationally expensive methods are applied. By eliminating structural data requirements, our approach dramatically expands the range of proteins that can be studied, including those resistant to crystallization or with highly flexible regions. Our comprehensive analysis provides deeper insights into protein-protein interactions than methods focused on single metrics.

Future work could incorporate graph-based representations to better model long-range dependencies. Additionally, expanding the training data to improve generalization and exploring hybrid approaches that combine sequence-based and structure-based methods can be beneficial. 

By providing accurate predictions without structural constraints, Twin Peaks opens new possibilities for studying protein function in systems previously inaccessible to computational methods, enabling novel applications in protein engineering, drug discovery, and understanding disease mechanisms.

\section*{Funding}
This work was partially funded by — Iowa State Startup Grant; Building A World of Difference Faculty Fellowship,
NSF EPSCoR RII Track-1, Award Number DQDBM7FGJPC5, and
Iowa Economic Development Authority Award Number: 24IEC006 to RC

\bibliographystyle{plain}
\bibliography{references}

\end{document}